\documentclass[prb,aps,floatfix]{revtex4}
\usepackage{graphicx,epsfig}
\begin{document}
\title{\bf Phonon-mediated negative differential conductance in molecular quantum dots}
\date{\today}

\author{Alex Zazunov}
\affiliation{Centre de Physique Th\'eorique, Case 907 Luminy, 13288 Marseille cedex 9, France}
\affiliation{Laboratoire d'Etudes des Propri\'etes
Electroniques des Solides, CNRS, BP 166, 38042 Grenoble, France}
\author{Denis Feinberg}
\affiliation{Laboratoire d'Etudes des Propri\'etes
Electroniques des Solides, CNRS, associated with Universit\'e Joseph Fourier, BP166, 38042 Grenoble, France}
\author{Thierry Martin}
\affiliation{Centre de Physique Th\'eorique and Universit\'e de la M\'editerann\'ee, 
Case 907 Luminy, 13288 Marseille cedex 9, France}

\begin{abstract}
Transport through a single molecular conductor is considered, showing negative 
differential conductance behavior associated with phonon-mediated electron tunneling processes. 
This theoretical work is motivated by a recent experiment by Leroy et al. using a carbon nanotube 
contacted by an STM tip [Nature {\bf 432}, 371 (2004)], where negative differential 
conductance of the breathing mode phonon side peaks could be observed. 
A peculiarity of this system is that the tunneling couplings which inject electrons 
and those which collect them on the substrate are highly asymmetrical. 
A quantum dot model is used, coupling a single electronic level to a local phonon,
forming polaron levels. A ``half-shuttle'' mechanism is also introduced.  
A quantum kinetic formulation allows to derive rate equations. 
Assuming asymmetric tunneling rates,
and in the absence of the half-shuttle coupling, negative differential 
conductance is obtained for a wide range of parameters. A detailed explanation of this 
phenomenon is provided, showing that NDC is maximal for intermediate electron-phonon coupling.  
In addition, in absence of a gate, the "floating" level 
results in two distinct lengths for the current plateaus, related to the 
capacitive couplings at the two junctions. It is shown that the "half-shuttle" mechanism tends 
to reinforce the negative differential regions, but it cannot trigger this behavior on its own.   
\end{abstract}

\maketitle

\section{Introduction}

The prospect of using molecules as the fundamental building blocks of future nanoelectronics devices is 
rather innovating and exciting from the point of view of potential applications. On the fundamental side, 
the field of molecular electronics also opens new directions for research because of the 
prominent role of phonon (vibron) excitations in electronic transport. The nano-objects which are connected to metallic 
leads may consists of individual molecules, self-assembled monolayers, or conjugated systems such as 
polymers and carbon nanotubes. The interplay of electron transport and molecular vibrations has triggered much 
interest, and unambiguous signatures of phonons have been detected in several experiments 
\cite{park,zhitenev,natelson,qiu,leroy,leroy2,pasupathy,sapmaz}. 
 
A recent work \cite{leroy,leroy2} considered electron 
injection from a scanning tunneling microscope (STM) tip into a carbon nanotube. 
A single-wall carbon nanotube (CNT) is freely suspended over a trench.
The STM-tip is located near the center of the suspended part of the CNT.
A DC-bias voltage $V$ is applied between the substrate and the STM, and 
the current flowing through the STM-tip - CNT - substrate structure is 
measured at a given tunneling distance, controlled by the 
setpoint current.   
The motivation for this special geometry 
is to allow for free internal vibrations to occur in the suspended portion of the nanotube, 
in particular the so-called
radial breathing modes (RBM) \cite{dresselhaus}. 
Contrary to acoustic modes in such system, this mode has a 
rather high oscillation frequency, and the authors observed that conductance peaks 
in the current-voltage characteristics of the nanotube were surrounded by 
phonon side peaks, due to emission or absorption of RBM phonons. Indeed, CNT display a sizeable coupling 
of electrons to RBM modes \cite{avouris}. The contacts to the nanotube 
being rather resistive, transport in this system is dominated by the Coulomb blockade regime, and phonon 
side peaks were observed around each Coulomb blockade peak in differential conductance
plots. 

Interestingly, the authors mention frequent detection of negative differential conductance (NDC) 
regions. Striking NDC features also appear in a very recent work by Sapmaz {\it et al.} \cite{sapmaz}, 
in a transport measurement of a suspended CNT, where current flows through the CNT, 
between two contacts at the substrate. In this work, 
phonon side peaks are attributed to longitudinal stretching modes, and the steps 
in the $I(V)$ characteristics are followed by spikes, thus displaying NDC features. 
The purpose of the present theoretical work is to show that such NDC features can be described 
quite simply using a generic model which consists of a quantum dot with a single orbital level, 
coupled to on-site single phonon mode, and connected to leads by tunnel junctions. 
Due to the weak tunnel couplings, the physics of NDC appear to be a consequence of the transport 
through small polaron states on the molecule. Notice that polaron formation in CNT 
is suggested by several works \cite{avouris}. 
Similar models have previously been considered in the 
literature \cite{schoeller,lundin,mccarthy,flensberg,aji,aleiner,vonoppen}. Here we consider 
a molecular system -- or a nanotube setup -- 
whose tunneling matrix elements from the molecule
to the leads are asymmetric, which is the case in Ref. \onlinecite{leroy}, and 
possibly also in Ref. \onlinecite{sapmaz}. 
We propose that given this asymmetry, NDC can be obtained for a wide 
range of parameters and it can even be quantified by analytical means for strong asymmetry. 
Furthermore, we argue that our approach for the description of NDC is by 
no means confined to the experimental geometry of Ref. \onlinecite{leroy}, 
which can be considered as an ``experimental paradigm''
for phonon modulated molecular transport. The present approach  should apply to 
any molecular transport in the Coulomb blockade
regime where an optical-like phonon mode dominates. Notice that NDC was 
found in Ref.\onlinecite{schoeller} within a symmetric model, 
using an ansatz for the Franck-Condon factors coupling the leads to the polaron levels. 
Catastrophic NDC was also proposed to occur in Ref. \onlinecite{mccarthy}, 
but due to an additional mechanism. 
In the framework of a single site model, NDC was also found 
in an adiabatic treatment of the polaron problem \cite{galperin}. It has been proposed 
to occur in the case of two competing molecular states \cite{nowack} 
or within a two-site model \cite{kaat}. Besides electron-phonon coupling, other physical 
mechanisms can lead to NDC \cite{hettler}.

In molecular electronics, one is tempted to use existing theoretical tools developed in the context 
of mesoscopic physics because the size of the devices allows a coherent description of the transport 
inside the molecule. When the molecule is connected to good contacts, 
and when the role of Coulomb interaction and electron-phonon interactions 
is reduced, the Landauer formulation of transport combined with a Green's 
function calculation of the transmission coefficient serves as a good 
starting point for computing transport \cite{datta}, but it then neglects
the dynamical degrees of freedom of the molecule. Phonons can be included 
in these approaches, analytically \cite{wingreen}, or 
using numerical calculations \cite{ness_fisher,emberly} based 
on the work of Ref. \onlinecite{bonca_trugman}.
Alternatively, the Landauer approach can be generalized to a situation where 
interactions are restricted to a finite region 
\cite{meir_wingreen,lundin,flensberg} using non-equilibrium 
Green's functions. 

On the other hand, molecular contacts or the contacts to a nanotube are often 
of poor quality. It is then reasonable to think of the molecule as a 
quantum dot, which is subject to the Coulomb blockade. This point of view 
was adopted by several authors 
\cite{schoeller,lundin,mccarthy,flensberg,aleiner,zhu}, and is confirmed by experiments.
One of the advantages of modecular electronics comes from the 
size of the individual nano-object which is connected to the electrodes. 
Even at relatively large temperatures (a few Kelvins), the dynamics within 
the molecule/nanotube is fully phase coherent. On the other hand, the 
tunneling rates to the reservoirs are typically small compared to temperature.  
One can then derive rate equations
for the electron population of the dot. Here our goal is to use the simplest model 
which can account for the physics of the experiment of Ref. \onlinecite{leroy}. 
We therefore derive a kinetic equation approach which can 
account for the observed NDC effect, treating the electron-phonon coupling non-perturbatively
while still allowing for an intuitive understanding of the physics at hand. 
Furthermore, consequences can be 
derived on the electron transport as well as on the non-equilibrium phonon population. 
In the present work the local mode is not coupled to any environmental degrees 
of freedom, like phonon modes or electron-hole excitations in the substrate. 
Although such coupling is probably relevant in some experiments, the very high 
phonon quality factors ($Q > 20000$) obtained in other works \cite{leroy} 
justify to neglect it as a first step.     

Experimentally, the addition of voltage gates on a molecular transport
setup is still challenging but it is 
possible \cite{dan_ralph,leroy2} to approach three metallic probes in a 
nanometer scale region. Yet, there is also a motivation to study a 
setup where the molecular levels are "floating" instead of being fixed by a gate.  
Besides the strong asymmetry of the tunneling rates, 
a specific feature of our work is that NDC can occur in such absence of gate voltage,
and that the ratio of the right/left capacitance plays an important role, as 
it dictates the location of the molecular levels. 

The structure of the paper is as follows. The model is introduced in Sec. II, 
and the derivation of the rate equations is provided in Sec. III. The general considerations
about current transport and phonon occupation numbers are discussed in Sec. IV.
Numerical results are presented in Sec. V, illustrating the role of the 
asymmetry of the capacitances, of the shift of the dot level, and of the so-called 
half-shuttle mechanism on the NDC. We conclude in Sec. VI.

\section{Theoretical model}

Although the potential relevance of our approach applies to the 
carbon nanotube experiments of Refs. \onlinecite{leroy,leroy2}, we refer
to the central region between the leads as the quantum dot.   
This quantum dot is weakly coupled to two metallic
electrodes by tunnel junctions. In the STM geometry of 
Refs. \onlinecite{leroy,leroy2}, 
the electrodes represent the STM-tip and the substrate,
to which we refer as the 'left' ($l$) and 'right' ($r$) electrodes, respectively. 
Each tunnel junction ($j = l,r$) is characterized by
a resistance ($R_j$) and a capacitance ($C_j$). 
While $R_r$ and $C_r$ are constant
for a given contacted nanotube sample, 
$R_l$ and $C_l$ are functions of the tip-tube separation. The capacitances $C_l$ 
and $C_r$ are sample-dependent \cite{leroy3}. For generality, one can add a gate voltage 
$V_g$ and a gate capacitance $C_g$.
\begin{figure}[h]
\epsfxsize 7cm
\centerline{\epsffile{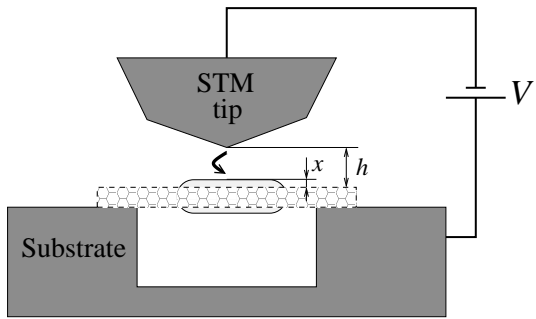}} 
\vspace{.5cm} 
\caption{Schematic drawing of a carbon nanotube suspended over a trench.
A bias voltage is applied between the STM tip (source) and substrate (drain).
The inflated portion in the suspended portion of the nanotube illustrates 
the radial breathing mode. $h$ is the tip--nanotube separation at rest.}
\label{setup}
\end{figure}

We focus on the strong Coulomb blockade regime, assuming that the number
of electrons which can be added to the dot is restricted to one. The physical features revealed 
in the corresponding bias window can be easily extended to a full span of many Coulomb 
blockade peaks, as well a several orbital levels as in Refs. \onlinecite{leroy,leroy2}.
The extra-charge electron state is locally coupled 
to a phonon mode: in Refs. \onlinecite{leroy,leroy2} this mode
is identified with the radial breathing mode (RBM) of the carbon 
nanotube \cite{dresselhaus,avouris} and, in Ref. \onlinecite{sapmaz}, with 
the longitudinal stretching mode. 
In view of the phonon-assisted tunneling processes,
the above-mentioned Coulomb blockade regime implies that 
the charging energy of the dot is assumed to be infinitely large
compared to the relevant energy scale determined (at low temperature) by 
the phonon energy $\Omega $ and the bias voltage. 
In the absence of phonons, electron transport would occur only through a 
single electron state on the dot.

In the Coulomb blockade regime, far from the Kondo regime,  
spin degrees of freedom are neglected.  
The Hamiltonian of the system is written as follows:
\begin{equation}
{\cal H} = {\cal H}_0 + {\cal H}_{leads} + {\cal V} \;,
\label{H}
\end{equation}
where
\begin{equation}
{\cal H}_0 = \left( \, \epsilon - g x \right) d^\dagger d + \Omega \, b^\dagger b  
\;,\;\;\; 
x = b + b^\dagger \;,
\end{equation}
\begin{equation}
{\cal H}_{leads} = \sum_{j k} \xi_{jk} \, c_{j k}^\dagger c_{j k} \;,
\end{equation}
\begin{equation}
{\cal V} = \sum_{j k} {\cal T}_j(x) \, c_{j k}^\dagger \, d + h.c. \;.
\label{V}
\end{equation}
Here the operator $d$ ($d^\dagger$) annihilates (creates) an electron  
on a single dot level of energy $\epsilon$; similarly, 
$c_{j k}$ ($c^\dagger_{j k}$) annihilates (creates) 
an electron with momentum $k$ and energy $\xi_{jk}$ in the $j$th lead.
The RBM is linearly coupled  (with the coupling energy $g$) 
to the electric charge on the dot; 
the RBM excitations are annihilated (created) by $b$ ($b^\dagger$). 
In Eq. (\ref{V}), ${\cal T}_j$ is the energy associated with the tunneling coupling to the dot;
for simplicity, ${\cal T}_j$ is assumed to be energy-independent
(constant density of states in the metallic leads), 
but we take into account the dependence of the tip-tube tunneling 
matrix amplitude on the boson coordinate $x$: due to 
the ``breathing'' motion of the tube, the tip-tube tunneling distance  deviates 
from its equilibrium value. Explicitly, we assume an exponential 
$x$-dependence:
\begin{equation}
{\cal T}_l(x) = {\cal T}_{l0} \, e^{- s x } \;,
\end{equation}
where $s$ is determined by the ratio of the amplitude of the zero-point RBM oscillations
to the electronic tunneling length ($\approx 0.5 \AA$) which characterizes the tunnel barrier 
between the STM tip and the nanotube.
Such position dependent amplitudes have been introduced in the 
context of nano-mechanical electronic devices\cite{gorelik,kotthaus,fedorets,nano_gorelik}, 
where one refers to the ``shuttle'' mechanism as the central region oscillates 
between the two electrodes.   
On the other hand, in our situation the tube-substrate tunneling matrix 
amplitude is $x$-independent: 
${\cal T}_r(x) \equiv {\cal T}_{r0}$. 
For $s\neq 0$, we refer here to this position dependent tunneling Hamiltonian
as the ``half-shuttle'' mechanism, because only one of the tunneling
amplitude (left) is modified by the position. 

In the model Hamiltonian (\ref{H}), formally describing the phonon-assisted 
tunneling of otherwise noninteracting electrons, 
the charging effects are taken into account via the bias-voltage dependence of 
the position of the  dot level ($\epsilon$) 
with respect to the chemical potentials of the leads, $\mu_{l,r}$. A gate voltage $V_g$ 
and gate capacitance $C_g$ can be included.
In our analysis, $\epsilon$ includes the change in the charging energy of the dot 
when one extra electron is added to the dot, and which is obtained 
from the electrostatic energy consideration: 
\begin{equation}
\epsilon - \epsilon_0 = 2E_C (n_x + 1/2) + e \phi~,
\label{energy_level}
\end{equation} 
where $\epsilon_0$ is the ``bare'' energy level of the electron level in the dot,
$E_C = e^2 / 2C$ is the charging energy and 
$e n_x$ is a background (fractional) charge of the dot.
Furthermore, the potential of the dot, $\phi$, is given by
\begin{equation}
\phi = \sum_{j = l,r} c_j \mu_j / e + c_g V_g~,
\label{phi_potential}
\end{equation} 
\begin{equation}
c_j = C_j /C~,~~~ c_g = C_g /C~,~~~
C = C_l + C_r + C_g~.
\end{equation} 
In typical experiments $C_g\ll C_j$. It is worth noticing that in the ``floating-level'' 
geometry which is considered in part of this work 
(e. g. without the gate electrode, $C_g=0$), the fractional charge is not fixed
but it can instead be affected by changes in 
the capacitances of the junctions \cite{Tinkham}   
(that is, $n_x$ may depend on the tip-tube separation). 

Assuming weak coupling to the leads,
it is convenient to eliminate the electron-phonon coupling in ${\cal H}_0$ by the unitary
transformation $\tilde{{\cal H}} = \Lambda^\dagger {\cal H} \Lambda$, with
\begin{equation}
\Lambda = e^{- i \alpha p \, d^\dagger d } \;,\;\;\; 
p = -i \left( b - b^\dagger \right) \;,\;\;\;
\alpha = g / \Omega  \;,
\end{equation}
In the rotated basis,
the electron state in the dot becomes ``dressed'' with phonons, forming a small polaron.
This results in the ``polaron shift'' of the dot level energy, 
$\tilde{\epsilon} = \epsilon - g^2/ \Omega$, and 
a renormalization of the dot-lead tunneling coupling. 
In the polaron representation, the Hamiltonian reads:
\begin{eqnarray}
\tilde{{\cal H}} &=& \tilde{{\cal H}}_0 + \tilde{{\cal V}} + {\cal H}_{leads}~,\\
\tilde{{\cal H}}_0 &=& \epsilon \, d^\dagger d + \Omega \, b^\dagger b\\
\tilde{{\cal V}} &=& \sum_{j k} {\cal T}_j(x) \, e^{- i \alpha p} \, c_{j k}^\dagger \, d + h.c. \;.
\label{tildeHt}
\end{eqnarray}
In Eq. (\ref{tildeHt}), we have used the fact that
\begin{equation}
\Lambda^\dagger  {\cal T}_l(x) \, d \, \Lambda = 
{\cal T}_l(x + 2 \alpha \, d^\dagger d ) \, d \, e^{- i \alpha p} =
{\cal T}_l(x) \, e^{- i \alpha p} \, d~.
\end{equation}

\section{Rate equations} 

Based on the assumption that the leads are in thermal equilibrium 
at given chemical potentials ($\mu_l$ and $\mu_r$), independently 
of the state of the dot, 
one can derive a kinetic equation for the reduced density matrix 
by tracing out the electrode degrees of freedom from the total density matrix. Such equations have been 
used by several authors,\cite{schoeller,mccarthy,aleiner,fedorets} but for sake of completeness 
we provide here a full derivation.

In the polaron representation, the reduced density matrix of the phonon-coupled dot is given by
\begin{equation}
R(t) = e^{-i t \tilde{{\cal H}}_0} R_V(t) \, e^{i t \tilde{{\cal H}}_0} \;,\;\;\;
R_V(t) = {\rm Tr}_{leads} \tilde{\rho}_V (t) \;,
\end{equation}
where $\tilde{\rho}_V(t)$ is the total density matrix in the interaction picture
(with respect to the tunneling coupling to the leads),
which obeys the operator equation of motion:
\begin{eqnarray}
i \partial_t \tilde{\rho}_V(t) = 
\left[ \, \tilde{{\cal V}}(t) , \tilde{\rho}_V(t) \right] \;,
\label{tilderhoV_eq}\\
\tilde{{\cal V}}(t) = 
e^{i t \left( \tilde{{\cal H}}_0 + {\cal H}_{leads} \right)} \, \tilde{{\cal V}} \, 
e^{-i t \left( \tilde{{\cal H}}_0 + {\cal H}_{leads} \right)}~.
\end{eqnarray}
Using the integral form of Eq. (\ref{tilderhoV_eq}),
\begin{equation}
\tilde{\rho}_V (t) = -i \int_0^t d \tau \, \left[ \, \tilde{{\cal V}}(\tau ) , 
\tilde{\rho}_V (\tau ) \right] + \tilde{\rho}_V (0)\;,
\label{tilderhoV_inteq}
\end{equation}
substituting Eq. (\ref{tilderhoV_inteq}) into Eq. (\ref{tilderhoV_eq}), 
and taking the trace over the electronic degrees of freedom of the leads
with the above mentioned assumption of 
$\tilde{\rho}_V(t) = \rho^{eq}_{leads} \otimes R_V(t)$,
where $\rho^{eq}_{leads}$ is the equilibrium density matrix of the leads,
we arrive at the integro-differential equation of motion for the
reduced density matrix in the polaron representation:
\begin{eqnarray}
\partial_t R_V(t) =
- \sum_{j=l,r} 2 \pi \nu_j \int_0^t d \tau \, e^{i \tilde{\epsilon} \tau} \, 
\left[ \, F_j^{>}(\tau) \, \left\{
\tilde{{\cal T}}_j^\dagger (t) \tilde{{\cal T}}_j(t-\tau ) \, d^\dagger d \, R_V(t - \tau) - 
\tilde{{\cal T}}_j(t-\tau ) \, d \, R_V(t - \tau)\, d^\dagger \tilde{{\cal T}}_j^\dagger (t)
\right\} \right. \nonumber \\ \left. + \;
F_j^{<}(\tau) \left\{ \, R_V(t - \tau) \, d d^\dagger
\tilde{{\cal T}}_j(t-\tau ) \tilde{{\cal T}}_j^\dagger (t) - 
\tilde{{\cal T}}_j^\dagger (t) \, d^\dagger \, R_V(t - \tau) \, d \, \tilde{{\cal T}}_j(t-\tau ) 
\right\} \right] + h.c. \,,
\label{kin_eq}
\end{eqnarray}
where $\nu_j$ is the density of states in the $j$th electrode,
\begin{equation}
\tilde{{\cal T}}_j(\tau ) = {\cal T}_j[(x(\tau )]  e^{- i \alpha p(\tau )}~,
\end{equation}
with 
\begin{eqnarray}
x(\tau ) & = & b \, e^{- i \Omega \tau } + h.c.~, \\
p(\tau ) & = & - i b \, e^{- i \Omega \tau } + h.c.~, 
\end{eqnarray}
and
\begin{equation}
F_j^{>,<}(\tau) = \frac{1}{2} \, \left[ \, \delta(\tau) \mp \, {\rm P.v.} \, 
\frac{i e^{- i \mu_j \tau }}{\beta \sinh \left( \pi \tau /\beta \right)} \right]
\label{FF}
\end{equation}
(where "P.v." stands for the principal value).
From the above expressions for the kernels $F_j^{>,<}(\tau )$,
it follows that the relevant retardation time is of the order of the inverse temperature,
$\beta$. Assuming temperature to be high compared to the tunneling rates,
$\beta^{-1} \gg 2 \pi \nu_j {\cal T}_j^2$, and considering the long-time behavior 
of the reduced density matrix, $t \gg \beta$, we construct an asymptotic solution
of Eq. (\ref{kin_eq}) by using an improved perturbation expansion \cite{zazunov}:
\begin{equation}
R_V(t) = \bar{R}(\lambda t) + \lambda \, R^{(1)}(t) + o(\lambda) ~.
\label{expansion}
\end{equation}
In Eq. (\ref{expansion}), $\lambda$ is a formal perturbation parameter 
that reflects a weakness of tunnel coupling 
and must be formally attributed to the right-hand side of Eq. (\ref{kin_eq}):
\begin{eqnarray}
\lambda \partial_z \bar{R}(z) + \lambda \, \partial_t R^{(1)}(t) + o(\lambda) =
- \lambda \sum_{j=l,r} 2 \pi \nu_j \int_0^t d \tau \, e^{i \tilde{\epsilon} \tau} \, 
\left[ \, F_j^{>}(\tau) \, \left\{
\tilde{{\cal T}}_j^\dagger (t) \tilde{{\cal T}}_j(t-\tau ) \, d^\dagger d \, 
\bar{R}(z - \lambda \tau) + ... \right. \right. ~, 
\label{kin_eq2}
\end{eqnarray}
with $z = \lambda t$. 
In the first-order equation in $\lambda$, 
retardation effects are neglected (the Markovian approximation), and in the above equation, 
we can replace $\int_0^t d \tau \, ... \,  \rightarrow \int_0^{+\infty} d \tau \, ... \;$.
By constraction,
$\bar{R}$ is, on the time scale $\beta$, a slowly evolving part of the reduced density matrix,
while the rapidly oscillating (with frequencies determined by a multiple integer of $\Omega$)
part of the right-hand side of Eq. (\ref{kin_eq2}) is absorbed into $R^{(1)}$. 
In what follows, we assume that the phonon frequency is high compared to the tunneling rates, 
$\Omega \gg 2 \pi \nu_j {\cal T}_j^2$. This ``anti-adiabatic'' condition allows us to neglect
$R^{(1)}$ as being negligibly small compared to $\bar{R}$.
Under the above assumptions, Eq. (\ref{kin_eq2}) reduces to 
a differential operator equation for $\bar{R}(t)$.

Although $\bar{R}$ is diagonal in the on-dot electron number
(no coherence between subsequent tunneling events at long times),
\begin{equation}
\bar{R}(t) = \left( 1 - d^\dagger d \right) R^{0}(t) + d^\dagger d \, R^{1}(t) \;,
\end{equation}
yet it is non-diagonal in the phonon number due to the presence of the displacement operators 
($e^{\pm i \alpha p}$) in the right-hand side of Eq. (\ref{kin_eq2}). 
However, equations for the diagonal and off-diagonal parts of $\bar{R}(t)$ are 
decoupled from each other. As a result, in steady state ($t \rightarrow +\infty$),
which we are interested in, $\bar{R}$ becomes diagonal in both the electron and phonon numbers.

Introducing the electron-phonon joint probabilities
\begin{equation}
P^{i}_n(t) = \langle n |R^{i}(t) | n \rangle~,
\end{equation}
for $i = \{0,1\}$ additional electrons and
$n = \{0, 1, 2, ...\}$ boson excitations on the dot,  
we obtain the system of rate equations for the joint probabilities,
which in steady case can be written in the following form:
\begin{eqnarray}
\partial_t P^{0}_n = 0 =
- \Gamma^{<}_n \, P^{0}_n + \sum_m \, L^{>}_{nm} \, P^{1}_m \;,
\nonumber \\
\partial_t P^{1}_n = 0 =
- \Gamma^{>}_n \, P^{1}_n + \sum_m \, L^{<}_{nm} \, P^{0}_m \;,
\label{rate_eqs}
\end{eqnarray}
with the normalization condition, $\sum_n \left( R^0_n + R^1_n \, \right) = 1$.
The charge/phonon transition rates are given by
\begin{equation}
\Gamma^{<}_n = \sum_m \, L^{<}_{mn} \;,\;\;\;
\Gamma^{>}_n = \sum_m \, L^{>}_{mn} \;,
\label{Gamma_n}
\end{equation}
\begin{equation}
L^{<}_{nm} = \sum_{j = l,r} \, \Gamma_j \, \gamma^2_{j,mn} \, 
f_j(\tilde{\epsilon} - \Omega_{mn}) 
\;,\;\;\;
L^{>}_{nm} = \sum_{j = l,r} \, \Gamma_j \, \gamma^2_{j,nm} \,
\left( 1 -  f_j(\tilde{\epsilon} - \Omega_{nm})\right) \;.
\label{L_ls_gr}
\end{equation}
In the above equations, 
$f_j(\omega ) = \left[ \, e^{\beta \left( \omega - \mu_j \right)} + 1 \right]^{-1}$ 
is the Fermi distribution function of the $j$th electrode,
$\Omega_{nm} = (n-m)\, \Omega$,
\begin{equation}
\Gamma_j = 2 \pi \nu_j \, {\cal T}_{j0}^2 \;,
\end{equation}
\begin{equation}
\gamma_{l,nm} = \langle n | e^{-s x } e^{-i \alpha p} | m \rangle \;,\;\;\;
\gamma_{r,nm} = \langle n | e^{-i \alpha p} | m \rangle \;.
\end{equation}
Explicitly, we have for the oscillator matrix elements:
\begin{equation}
\langle n | e^{-s x } e^{-i \alpha p} | m \rangle =
e^{-s \alpha  + {1 \over 2} \left( s^2 - \alpha^2 \right)} \, 
\sum_{q=0}^{{\rm Min}(n,m)} \, \frac{(-1)^{m-q} \sqrt{n! m!}}{(n-q)! (m-q)! q!} \; 
\left( \alpha - s \right)^{n-q} \left( \alpha + s \right)^{m-q}  \;.
\end{equation}

From the solution of Eq. (\ref{rate_eqs}) we can calculate
the DC current and expectation values of the phonon observables.
It should be noticed that in the original (non-polaron) representation, 
the resulting phonon density matrix is given by
\begin{equation}
R_{ph} = \sum_n \, \left( P^{0}_n \, | n \rangle \langle n | +     
P^{1}_n \, e^{-i \alpha p} \, | n \rangle \langle n | \, e^{i \alpha p} \right) \,.
\label{ph_distrib}
\end{equation}
In general, $R_{ph}$ has contributions from both charge states $i = 0,1$ 
and is non-diagonal in vibrational space (two shifted oscillators). 

In order to express the current in terms of the joint probabilities $P^i_n$, 
we have to average the current operator taken in the polaron representation,
\begin{equation}
\tilde{I}_j = e i \sum_k {\cal T}_j(x) \, e^{- i \alpha p} \, c_{j k}^\dagger \, d + h.c. \;,
\end{equation}
with the total density matrix 
\begin{equation}
\tilde{\rho}(t) = 
e^{-i t \left( \tilde{{\cal H}}_{dot} + {\cal H}_{leads} \right)} \, 
\tilde{\rho}_V(t) \, 
e^{i t \left( \tilde{{\cal H}}_{dot} + {\cal H}_{leads} \right)}~,
\end{equation}
where $\tilde{\rho}_V(t)$ obeys Eq. (\ref{tilderhoV_inteq}). 
Following the same steps and assumptions which we have made 
in the derivation of the rate equations, 
we obtain the average DC current flowing from the $j$th lead to the dot as (we set $e = 1$)
\begin{equation}
I_j =  \sum_n \, 
\left( \Gamma^{<}_{j,n} \, P^{0}_n - \Gamma^{>}_{j,n} \, P^{1}_n \right) \,,
\label{Ij}
\end{equation}
where $\Gamma^{<,>}_{j,n}$ are the partial relaxation rates 
contributing to $\Gamma^{<,>}_n$,
Eq. (\ref{Gamma_n}), from the $j$th electrode:
\begin{eqnarray}
\Gamma^{<}_{j,n} & = & \sum_m \, \Gamma_j \, \gamma^2_{j,nm} f_j(\tilde{\epsilon} - \Omega_{nm})~, \\ 
\Gamma^{>}_{j,n} & = & \sum_m \, \Gamma_j \, \gamma^2_{j,mn} \, \left( 1 -  f_j(\tilde{\epsilon} - \Omega_{mn})\right)~.
\end{eqnarray}
By virtue of the rate equations, we have conservation of the current, $I_l = - I_r$.

\section{Phonon-assisted transport for $\Gamma_l \ll \Gamma_r$} 

In calculating $I(V)$-characteristics from Eqs. (\ref{kin_eq}) and (\ref{Ij}),
we will focus on the case of highly asymmetric two-junction model with 
$\Gamma_l \ll \Gamma_r$. This corresponds to the typical experimental 
situation with STM measurements where a contact to the STM tip
plays the role of the high resistive tunneling junction.
For instance, in STM measurements on suspended nanotubes, 
the typical ratio $R_{tip} /R_{sub} \sim 10^3-10^5$ can be huge 
depending, in particular, on the tunneling distance
between the STM-tip and the nanotube, $h \sim 4 \; \AA$ typically. In the 
present Section, no gate is present (floating level) thus $C_g=0$.

At the same time, according to the data in Ref. \onlinecite{leroy2} 
obtained from spectroscopy measurements on suspended nanotubes, 
the ratio $C_{tip} / C_{sub} \equiv C_l / C_r$ can be smaller as well as larger than unity
depending on an effective length of the portion of the nanotube that is 
on the substrate, and which can be relatively short due to local defects induced 
by the edges of the trench. Thus, there is no dominating asymmetry in 
the capacitive coupling of the dot to the both electrodes. As a result,
in the voltage-biased system, the effective position of the polaron level 
with respect to the chemical potentials of the leads is strongly affected by 
the ratio $C_l / C_r$ which is not negligibly small like $\Gamma_l / \Gamma_r$, 
or can even be larger than unity \cite{leroy3}. It results that the polaron level 
(assuming neither gate-induced nor intrinsic shift of the polaron level 
at zero voltage) is not trivially stuck to the chemical potential of the far less 
resistive junction electrode, $\mu_r$.

Considering the system of stationary rate equations, Eq. (\ref{rate_eqs}), 
and writing explicitly expressions for the Fermi factors entering 
Eq. (\ref{L_ls_gr}), we obtain 
[see also Eqs. (\ref{energy_level}), (\ref{phi_potential})]:
\begin{equation}
f_l(\tilde{\epsilon}) = f(E - c_r V)  \;,\;\;\;
f_r(\tilde{\epsilon}) = f(E + c_l V)   \;,\;\;\; V = \mu_l - \mu_r \;,
\label{f_lr}
\end{equation}
where $f(\omega ) = \left( \, e^{\beta \omega} + 1 \right)^{-1}$, and
\begin{equation}
E = \epsilon_0 - g^2 / \Omega + 2 E_C (n_x + 1/2)
\end{equation} 
determines the position of the polaron level with respect to the lead chemical potentials at $V=0$,
considered as the reference level of zero energy.
From Eq. (\ref{f_lr}), the role of charging effects 
in the single-electron resonant level problem can be viewed as follows : 
when changing the bias voltage, the chemical potentials move in opposite directions
(depending on the sign of $V$) with different ``velocities'' determined 
by the capacitance ratios, $c_r$ and $c_l$ for the left and right electrode, respectively. 
The above picture is given in the "reference frame" of the polaron level,
where its position is voltage-independent. 
For the floating-dot geometry, 
$E$ may be strongly affected by a background charge, $n_x$. 
In the presence of a gate electrode \cite{leroy3}, $E$ can be controlled 
by the gate potential, and $C$ in Eq. (\ref{phi_potential}), 
must also include the gate capacitance, with $c_l + c_r < 1$.
 
In this section, in order to make analytical progress on the prediction of the current transport, 
we formally assume temperature to be zero.
Because the rate equations have been derived in the high-temperature approximation
($T \gg \Gamma_r$), the zero-temperature assumption is justified for the voltage-biased cases
where the chemical potentials, $\mu_{l,r}$, are not very close to the phonon sidebands,
so that thermally activated tunneling processes can be neglected.

We first notice that in the case where the polaron level $E$ lies 
beyond the bias-voltage window, the system of rate equations, 
Eq. (\ref{rate_eqs}), reduces to the following: 
\begin{equation}
L^{<}_{nm} \, P^{0}_m = L^{>}_{mn} \, P^{1}_n \;,
\label{det_ball}
\end{equation} 
which is satisfied for all phonon numbers $m,n$. 
The ground state solution of Eq. (\ref{det_ball}) is either 
$P_n^i=\delta_{n0} \delta_{i 1}$ ($i=0,1$) for $E < \mu_{l,r}$,
or $P_n^i=\delta_{n0} \delta_{i 0}$ for $E > \mu_{l,r}$.
As a result, in this case the current is zero :
no contribution to the current occurs from the phonon sidebands alone
(cf. Ref. \onlinecite{aleiner}).

Eq. (\ref{det_ball}) also works in the less trivial case
when $E$ lies within the bias-voltage window, but $|\mu_{l,r} - E| < \Omega$. 
For instance, for $V \equiv \mu_l - \mu_r > 0$, 
the solution of Eq. (\ref{det_ball}) (valid for arbitrary $\Gamma_{l,r}$)   
reads $P_n^i=\delta_{n0} P^i$, where 
\begin{eqnarray}
P^{0} = 1 - P^1 &=& \frac{L^{<}_{00}}{L^{<}_{00} + L^{>}_{00}} = 
\frac{\tilde{\Gamma}_l}{\tilde{\Gamma}_l + \Gamma_r}~, \\
\tilde{\Gamma}_l &=& \Gamma_l e^{- 2 s \alpha + s^2} \;,
\end{eqnarray} 
and the current is given by 
\begin{equation}
I = e^{-\alpha^2} \, 
\frac{\tilde{\Gamma}_l \Gamma_r}{\tilde{\Gamma}_l + \Gamma_r} \;.
\label{curr_through_polaron}
\end{equation} 
For large $\alpha$, the exponential prefactor in Eq. (\ref{curr_through_polaron}) leads 
to a significant suppression of the current at low bias voltages 
(the so-called Franck-Condon blockade \cite{flensberg,aleiner,vonoppen}). 
The role of $s$ (``half-shuttle'') will be discussed later.

At higher voltages, such that phonon sidebands enter the bias-voltage window, 
Eq. (\ref{det_ball}) fails, and we have to go back to the full system of rate equations, 
Eq. (\ref{rate_eqs}), taking into account the total balance 
of probabilities for transitions between different states of the dot.
This renders the problem of calculating phonon distribution functions, $P^i_n$, 
and eventually the $I(V)$-characteristics, a rather complicated one 
to be solved analytically. 
However, for the case of high asymmetry in the tunneling couplings,
it is possible to make some further analytical progress before passing 
to numerical calculations \cite{Tinkham}.

\begin{figure}[h]
\epsfxsize 6cm
\centerline{\epsffile{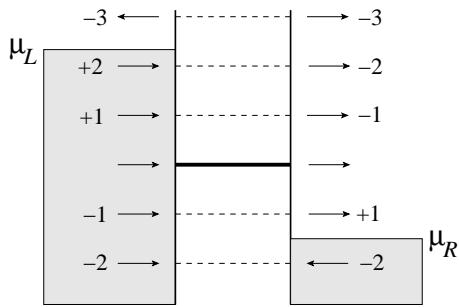}} 
\vspace{.5cm} 
\caption{Energy-level diagram for the case $M=1$ and $N=2$; 
the arrows show possible channels for electrons to tunnel onto/from the dot
with changing (indicated by numbers) the phonon occupancy; $\mu_l-\mu_r=V$}
\label{energy_dgrm}
\end{figure}

We now exploit the condition $\Gamma_l \ll \Gamma_r$ for analyzing 
steady solutions of the rate equations. 
In the general case, when a bias voltage $V$ is applied and 
the current is not identically zero (up to thermally activated processes),
the polaron level is accompanied by $N + M$ phonon sidebands captured in 
the bias-voltage window,
where we introduce (``int'' stands for the integer part)
\begin{eqnarray}
N&=& {\rm int}(|\mu_l-E| / \Omega) \label{NNN}~,\\ 
M&=& {\rm int}(|\mu_r-E| / \Omega) \;.\label{MMM}
\end{eqnarray} 
Independently of the sign of $V$, 
the integer $N$ ($M$) is defined as a number of phonon sidebands lying between the polaron level
and the chemical potential of the electrode connected to the dot 
through the more (less) resistive junction. 
An example with $N=2$ and $M=1$ for $V>0$ is shown on Fig. \ref{energy_dgrm}.  
Due to high asymmetry $\Gamma_l \ll \Gamma_r$,
only probabilities $\{ P^0_m \}$ with 
$m \in [0, M]$ do not vanish in this limit.
For the example on Fig. \ref{energy_dgrm}, the probabilities to have an electron on the dot 
with any $n$ phonons, $P^1_n$, are suppressed 
due to "fast" tunneling of the electron to the right electrode via $n+2$ open channels; 
the probability to have the dot with an empty electron state 
but with the number of phonons $M + 1 =2$ (and higher), $P^0_{M+1}$, 
is also negligible due to the ``fast'' tunneling of an electron from the right electrode 
with absorbing two phonons. 
In other words, on a large time scale determined by $\Gamma_l^{-1}$,
the polaron-hole states with 0 and 1 phonon excitation are 
(quasi)steady states with respect to the tunneling coupling to the right electrode. Notice 
that in the asymmetric situation, the phonon number distributions are nearly the same in 
the polaron basis (the $P_n^0$'s) and the original basis ($R_{ph}$), 
as shown by Eq. (\ref{ph_distrib}).

The case $V<0$ can be treated in the same manner.
We obtain that only probabilities $\{ P^1_m \}$ with 
$m \in [0, M]$ are not vanishing as $\Gamma_l / \Gamma_r$.
As a result, with a good accuracy, controlled by the smallness of $\Gamma_l / \Gamma_r$,
the current flowing from the left to the right, 
$ I \equiv I_l$ [see Eq. (\ref{Ij})], can be written as  
\begin{equation}
I(V>0) = \sum_{m=0}^M P^0_m \, I_m \;,\;\;\;
I_m = \Gamma_l \sum_{n=0}^{m+N} \gamma^2_{l,mn} \,,
\label{I_posV}
\end{equation} 
where $\sum_{m=0}^M P^0_m + O(\Gamma_l / \Gamma_r)  = 1 $, and
\begin{equation}
I(V<0) = - \sum_{m=0}^M P^1_m \, I_m \;,\;\;\;
I_m = \Gamma_l \sum_{n=0}^{m+N} \gamma^2_{l,nm} \,,
\label{I_negV}
\end{equation}
with $\sum_{m=0}^M P^1_m + O(\Gamma_l / \Gamma_r) = 1$.
The current is given by a sum of partial currents, $I_m$, 
representing $M$ conducting channels which are distributed with 
the corresponding phonon occupation probabilities. 
The magnitude of each partial current depends on the previously defined 
number of phonon sidebands $N$ captured between $E$ and $\mu_l$.

When varying $V$, the number of captured phonon sidebands, $N$ and $M$, can change.
In the voltage ranges where $N$ and $M$ do not change, we obtain plateaus in the $I(V)$ characteristics. 
Assuming for concreteness $V>0$, we see from Eq. (\ref{I_posV}) that 
if $V$ increases in such a way that only $N$ changes 
($N$ can only increase with increasing $V$), 
the magnitude of each partial current, and hence the net current, will increase.  
Thus, we obtain here the positive differential conductance (PDC) behavior
which is commonly computed and observed. 
On the contrary, if when increasing $V$, only $M$ changes (i.e. increases),
while $N$ remains constant, then a new phonon-assisted channel is added. 
This leads to redistribution of phonon occupation probabilities between all open channels.
Paradoxically, the net current may decrease, leading to 
a NDC behavior on $I(V)$ \cite{schoeller} .
  
For $V>0$, assuming that due to the voltage increase $V \rightarrow V + \Delta V$ 
the number of phonon sidebands in the bias-voltage window has changed as 
$\{N=0, M=0 \} \rightarrow \{N=0, M=1\}$, using Eq. (\ref{I_posV}) 
the variation of the current $\Delta I = I(V + \Delta V) - I(V)$ becomes:  
\begin{equation}
\Delta I = P^0_0 I_0 + P^0_1 I_1 - I_0 = 
- P_1^0 \, \tilde{\Gamma}_l e^{- \alpha^2} \, 
\left[ \, 1 - \left( \alpha - s \right)^2 - \left( 1 - \alpha^2 + s^2 \right)^2
\right] \,.
\label{deltaI}
\end{equation}
Thus, for this particular situation, assuming for simplicity $s =0$, 
the condition for the NDC is $\alpha < 1$: large values of the electron-phonon
coupling do not favor NDC.
The magnitude of
the negative step on $I(V)$ increases with increasing phonon occupancy $P_1^0$. 

For $V<0$, considering the same transition, $\{N=0, M=0\} \rightarrow \{ N=0, M=1 \}$, one finds:
\begin{equation}
\Delta I = - P_1^1 \, \tilde{\Gamma}_l e^{- \alpha^2} \, 
\left[ \, 1 - \left( \alpha + s \right)^2 - \left( 1 - \alpha^2 + s^2 \right)^2
\right] \,,
\label{deltaI_negV}
\end{equation}
with the same NDC condition, $\alpha < 1$, for $s=0$. 
From the above expressions, we also notice that for positive (negative) bias voltage and non-zero
but small half-shuttle $s \ll 1$, the NDC effect 
increases (decreases) for the first phonon-assisted step of the $I(V)$ characteristic. 

As far as the bias voltage increases, 
the current eventually saturates and does not change in practice.
For large bias voltages $V \rightarrow \pm \infty$,
the corresponding saturation currents $I_{sat}^{(\pm)}$ 
are given by (up to corrections $O(\Gamma_l^2 / \Gamma_r)$):
\begin{eqnarray}
I_{sat}^{(+)} & = &  
\Gamma_l \sum_{m=0}^{+\infty} P^0_m \, \langle m | e^{- 2 s x} | m \rangle~, \\
I_{sat}^{(-)} & = &  
- \Gamma_l \, e^{-4 s \alpha} \, 
\sum_{m=0}^{+\infty} P^1_m \, \langle m | e^{- 2 s x} | m \rangle \;.
\end{eqnarray}
In the absence of the half-shuttle mechanism, we obtain that the saturation current 
does not depend on the phonon distribution, 
$I_{sat}^{(\pm)} = \Gamma_l$. 
For $s \neq 0$, considering equilibrated phonons, 
i.e., forcing $P^s_m \rightarrow \delta_{m0}$ in the limit of strong phonon relaxation, 
we obtain the saturation currents:
\begin{equation}
I_{sat}^{(+)} = \Gamma_l \, e^{2 s^2} \;,\;\;\; 
I_{sat}^{(-)} = - e^{-4 s \alpha} \, I_{sat}^{(+)} \,.
\label{Isat_shuttle}
\end{equation}
For non-equilibrated phonons, Eq. (\ref{Isat_shuttle}) is expected 
to be accurate in the limit $c_l \ll c_r$. Nevertheless, for relatively small $\alpha < 1$,
the asymmetry in the current saturation values due to the half-shuttle
can be noticeable at low bias voltages for $c_l \sim c_r$ 
because of the fast saturation of the partial currents $I_m$
(fast convergency of the series in Eqs. (\ref{I_posV}) and (\ref{I_negV})).

\section{Numerical results and discussion} 

We now turn to the numerical solution of the rate equations, and explore the 
parameter space in order to find the signatures of NDC. 
After writting the system of the rate equations, Eq. (\ref{rate_eqs}), in the matrix form,
$\hat{\Lambda} \, P = 0$, with $P^T = \left( P^0_0, P^1_0, P^0_1, P^1_1, ... \right)$,
the problem reduces to finding the zero eigenvalue of the matrix $\hat{\Lambda}$.\cite{aleiner}  
The typical maximum number of phonon states has been taken around 40, 
so that results of the calculation would not depend on this choice 
for the considered range of parameters. 
From numerically found solutions $P$,
the current has been calculated in the left and right electrodes using Eq. (\ref{Ij}). 
The achieved tolerance for $I_l = - I_r$ was $10^{-6}$. 

Contrary to the previous analytical arguments, calculations are performed at finite temperature. 
For calculation purposes we assume a large asymmetry in the tunneling rates, 
$\Gamma_l / \Gamma_r = 10^{-4}$. Yet, most of our results also hold for moderate asymmetries. 
The current $I$ is plotted in units of 
$\Gamma_l \Gamma_r/(\Gamma_l + \Gamma_r)$, 
the bias voltage $V$, the polaron level $E$ and temperature $T$ are given in units of $\Omega$. 
In subsection A, one considers the floating level case. We consider 
the effect of a given shift of the dot level (due to background or gate charges) in subsection B. 
In these two subsections we will disregard the half-shuttle ($s = 0$), 
which will be considered in the last subsection.

\subsection{Non-shifted polaron level ($E=0$)}

We start by considering the case where 
the capacitances surrounding the dot fully specify the position of the polaron level.
At zero bias voltage,
the polaron level is aligned with the chemical potentials of the electrodes, $E=0$. 
In this case, as it immediately follows from the system of rate equations, 
Eq. (\ref{rate_eqs}),
we have the symmetry relation $P^0_n(V)=P^1_n(-V)$ and consequently $I(-V) = -I(V)$.
Notice that for the "floating-level" geometry the case of $E=0$ is an exceptional one,
and must be viewed as a reference point. 

\begin{figure}[!h]
\epsfxsize 7cm
\centerline{\epsffile{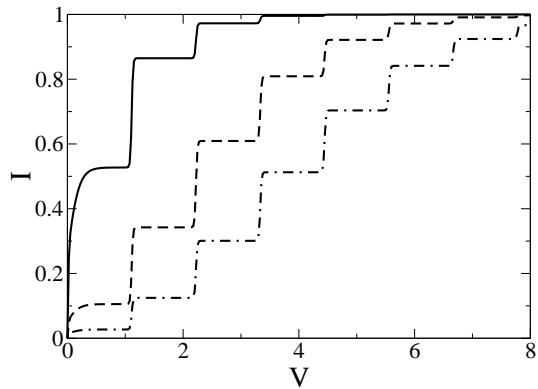}} 
\caption{
The case $c_l=0.1$ for $\alpha = 0.8$ (solid), 1.5 (dashed) and 1.9 (dot-dashed).
$T=0.01$.}
\label{ivPDC}
\end{figure}

Fig. \ref{ivPDC} shows the $I(V)$ characteristics for different $\alpha$'s assuming small $c_l =0.1$,
so that the polaron level lies closer to $\mu_r$. In the experiment of Refs. \onlinecite{leroy,leroy3},
this means that the nanotube has a large overlap with the conducting substrate.  
As long as $M=0$, that is if $V < \Omega / c_l$ 
(this condition is satisfied on the bias-voltage range which is plotted), 
$P^0_n = \delta_{n0} - O(\Gamma_l / \Gamma_r)$, which results in the PDC behavior of $I(V)$.
The PDC steps correspond to $N$ increasing  by one each time
$V$ passes through a multiple integer of $\Omega / c_r$. 
Explicitly, the current-step amplitude at $V = n \Omega / c_r$ corresponding to the current increase 
is given by 
\begin{equation}
\Delta I^{(n)} = \Gamma_l \, e^{-\alpha^2} \, \frac{\alpha^{2 n}}{n!} \;.
\end{equation}
For smaller $\alpha$'s the current-step amplitude decreases faster with increasing $V$,
which leads to the saturation of the current at lower voltages. 

Note that the first step (low bias voltage) is rounded. This feature is specific to the 
fact that the capacitances are asymmetric. Indeed, increasing the bias voltage from zero, 
the chemical potentials move away from the polaron level, but the chemical potential
whose lead has the largest capacitance remains close to this level. In this event, thermally 
activated tunneling processes can be effective. When the bias voltage is further increased
the phonon sidebands which contribute to new steps are those {\it above} the polaron level, 
and thus their corresponding steps are not thermally rounded. 
In the case where the asymmetry in the capacitances is opposite, which is shown 
in the next figure, the same reasoning applies. 

\begin{figure}[!!h]
\epsfxsize 7cm
\centerline{\epsffile{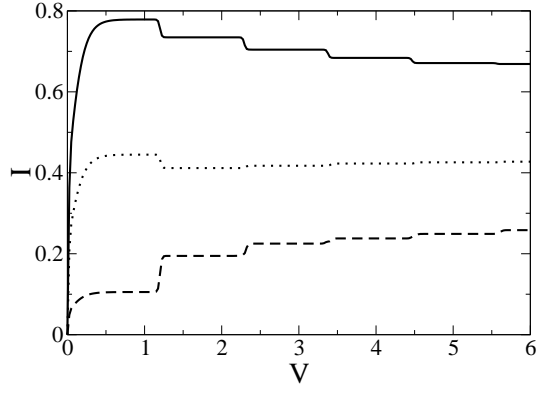}} 
\caption{
The case $c_l=0.9$ for $\alpha = 0.5$ (solid), 0.9 (dotted) and 1.5 (dashed).
$T=0.01$.} 
\label{ivNDC}
\end{figure}

Fig. \ref{ivNDC} shows the  $I(V)$ characteristics for different $\alpha$'s assuming
a relatively large $c_l =0.9$. In the experiment of Refs. \onlinecite{leroy,leroy3},
this means that the nanotube has a small length, thus a small overlap  with the 
conducting substrate, and thus a small
capacitance compared to that of the tip-nanotube contact. Alternatively, a large $c_l$
could be achieved if the nanotube contains impurities or bends.   

In the bias-voltage window, the polaron level is closer to $\mu_l$ than to $\mu_r$.
Within the bias-voltage range which is plotted,
the current changes by steps when $V$ passes through a multiple integer of $\Omega / c_l$, as before.
At these points, $M$ increases by one; correspondingly, one more phonon-assisted channel is added.
For $\alpha = 1.5$ one still observes the PDC behavior, although the height of the current steps
is strongly suppressed compared to the previous case on Fig. \ref{ivPDC}.
For $\alpha < 1$ NDC occurs in the first step, and with decreasing $\alpha$, 
more NDC steps appear. For all curves, 
Eq. (\ref{deltaI}) gives the current jump between the second
and first plateaus on Fig. \ref{ivNDC}.
We have also performed calculations of the current voltage characteristics in this NDC
regime while increasing gradually the temperature (not shown). Temperature tends to smooth
the steps and to suppress NDC altogether when it becomes comparable to the 
spacing between the steps. 

\begin{figure}[!h]
\epsfxsize 6cm
\centerline{\epsffile{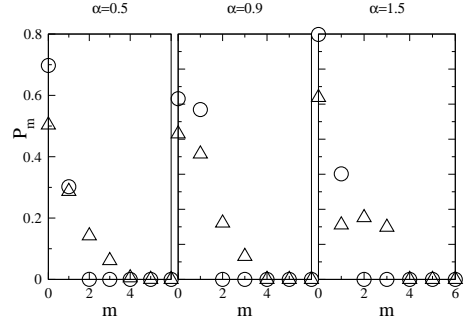}} 
\caption{Phonon distribution, $P_m^0$, for the case of Fig. \ref{ivNDC} 
at the bias voltage $V=2$ (circles) and $V=4$ (triangles).}
\label{pnNDC}
\end{figure}

At the same time that we monitor the electronic current, it is instructive to 
examine the phonon occupation numbers in order to quantify PDC or NDC behavior.  
The phonon distributions $P_m^0$, which play the role of probabilities of open channels, 
at different bias voltages are shown on Fig. \ref{pnNDC}. 
This figure demonstrates the increase of the number of phonon excitations 
with increasing $M$ (See Eq. (\ref{MMM})). It shows that the phonon number 
which are occupied is restricted by $M-1$. Thus, the phonon distribution allows 
to understand the height of the PDC or the NDC steps.

\begin{figure}[!h]
\epsfxsize 7cm
\centerline{\epsffile{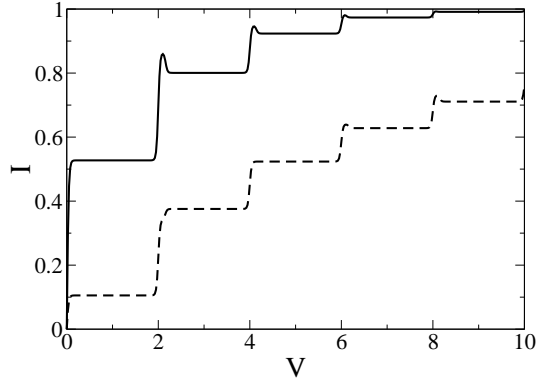}} 
\vspace{.5cm} 
\caption{The case $c_l=0.5$ for $\alpha = 0.8$ (solid) and 1.5 (dashed). $T=0.01$. }
\label{iv_cl05E0}
\end{figure}

\begin{figure}[!h]
\epsfxsize 5.5cm
\centerline{\epsffile{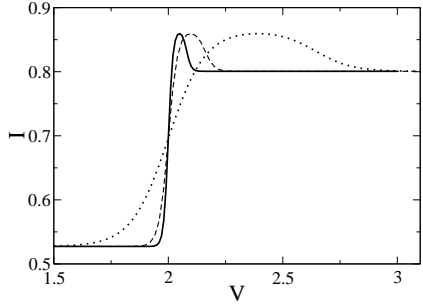}} 
\vspace{.0cm} 
\caption{Temperature dependence of the NDC-singularity 
for $\alpha = 0.8$: $T = 0.005$ (solid), 0.01 (dashed) and 0.04 (dotted).}
\label{smallpeak_Tdependence}
\end{figure}

In the two previous cases, two limits were considered : either $c_l \ll c_r$ or $c_l \gg c_r$. 
On the way to the more general situation when $c_l$ and $c_r$ are comparable,
let us now consider in more detail the capacitively symmetric case when $c_l= c_r = 0.5$.
The $I(V)$ characteristics for $\alpha = 0.8$ and 1.5 are shown on Fig. \ref{iv_cl05E0}.
For $\alpha = 1.5$, the current shows a stepwise increased with essentially no NDC behavior.  
Note that because the capacitances are equilibrated, the first step does not display substantial 
rounding due to thermal effects. 
For $\alpha = 0.8$, one observes a stepwise increase of the current, but the onset of each
step is associated with a small spike, thus exhibiting NDC behavior. This NDC behavior is
thus qualitatively different from the one previously observed for the case of asymmetric 
capacitances. 
The spikes bringing the NDC features to the $I(V)$ characteristics are robust with respect to temperature.
In Fig. \ref{smallpeak_Tdependence} one sees that these spikes are broadened with temperature, but NDC
persists as long as the temperature is not comparable to the phonon frequency. 
On the $I(V)$ for $\alpha = 1.5$ (Fig. \ref{iv_cl05E0}), 
the NDC-singularities appear only for higher voltages, however, the first phonon-assisted current step 
(at $V= 2 \Omega$) is also deformed. 

\begin{figure}[h]
\epsfxsize 7cm
\centerline{\epsffile{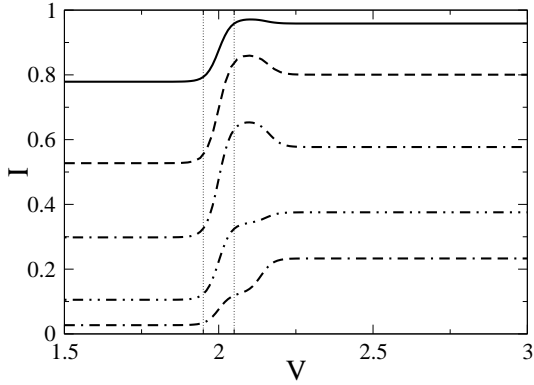}} 
\vspace{.0cm} 
\caption{From up to down $\alpha=\{ 0.5, 0.8, 1.1, 1.5, 1.9 \}$ 
for $c_l=0.5$. $T=0.01$.
The dotted vertical lines around $V=2$ (corresponding to $V_{\pm}=2 \pm 0.05$) 
show the width of the Fermi distribution function at given $T$.}
\label{iv_peaks}
\end{figure}

At low temperature, it is interesting to make a zoom of a given step in order to see how the spikes
evolve when the electron phonon coupling is varied. 
Fig. \ref{iv_peaks} shows these ``singularities'' on the second step of the  $I(V)$ characteristic 
for a wide  range of $\alpha$'s (cf. Fig. \ref{iv_cl05E0}). 
Note that the NDC behavior is more pronounced with increasing 
$\alpha = \{ 0.5, 0.8, 1.1 \}$. Indeed, in the limit of vanishing
electron phonon coupling, one gets only one step due to the polaron (electron)
level, and all NDC features are absent. There is therefore an optimal value of $\alpha$
which displays maximal NDC behavior.  Further increasing $\alpha$, NDC converts 
to the PDC case (the curves for $\alpha = \{ 1.5, 1.9 \}$).

In this case of equal capacitances ($c_l=c_r$), the polaron level lies in the middle of the bias-voltage 
window at all $V$, so that both $N$ and $M$ increase by one when $V$ passes through a multiple integer 
of $2 \Omega$. In particular, at $V=2 \Omega$ the transition $\{N=M=0\} \rightarrow \{N=M=1\}$ occurs : 
below $V=2 \Omega$ the polaron level is the only one which lies 
within the bias voltage window, while immediately above  $V=2 \Omega$, two phonon sidebands 
are simultaneously captured by the bias window. 
As a result, in the vicinity of $V=2 \Omega$ (how close depends precisely on temperature), 
we have a competition between the PDC- and "possible NDC"-type contributions to the current,
corresponding to the processes of electron tunneling from the left electrode into the dot 
with emission and absorption of one phonon, respectively.

In the transition region, $V \approx 2 \Omega$, the current is given by 
(here we take into account the Fermi factors)
\begin{equation}
I(V) = \sum_{m=0}^1 P^0_m(V) \, I_m ~,
\end{equation} 
\begin{equation}
I_m = \Gamma_l \sum_{n=0}^{m+1} \gamma^2_{l,mn} f(-c_r V - \Omega_{mn})~,
\end{equation}
and $P^0_0 + P^0_1 + O(\Gamma_l / \Gamma_r)  = 1$. Explicitly, we have
\begin{equation}
I(V) = \Gamma_l e^{-\alpha^2}  \left[ \, 1 + \alpha^2 f(\Omega - V/2) + 
P^0_1(V) \, \alpha^2 \left( 
-1 + \alpha^2 + \frac{1}{2} \left[
(1 - \alpha^2) (2 - \alpha^2) - \alpha^2 
\right] f(\Omega - V/2)
\right)
\right] \,.
\end{equation} 
>From the numerics, it follows that up to $V=V_{+}\equiv 2\Omega+W > 0$, where $W$ 
is the half-width of the Fermi distribution function (see caption of Fig. \ref{iv_peaks}), 
we have $P^0_1(V)=0$, thus the current successively increases up to: 
\begin{equation}
I(V_{+}) = \Gamma_l e^{-\alpha^2}  \left( \, 1 + \alpha^2 \right) \;.
\end{equation} 
This effect is due to the {\it high asymmetry} in the tunneling rates $\Gamma_{l,r}$: 
the probability to have an empty dot with one phonon, $P^0_1$,
remains negligible, $O(\Gamma_l/\Gamma_r)$, 
until there is some probability 
to have thermally activated electrons of energy $E-\Omega$ coming from the right electrode. 
Only then, for $V > V_{+}$, where $f(\Omega - V/2)$ has practically reached unity, 
$P^0_1(V)$ starts deviating from zero. Thus, we have
\begin{equation}
I(V) - I(V_{+}) = P^0_1(V) \, \Gamma_l e^{-\alpha^2} g(\alpha) \;,\;\;\; V > V_{+} \;,
\label{curr_diff}
\end{equation} 
\begin{equation}
g(\alpha) = 
\alpha^2
\left( 
-1 + \alpha^2 + \frac{1}{2} \left[
(1 - \alpha^2) (2 - \alpha^2) - \alpha^2 
\right]
\right) \;.
\label{g_vs_alpha}
\end{equation} 
\begin{figure}[!h]
\epsfxsize 5cm
\centerline{\epsffile{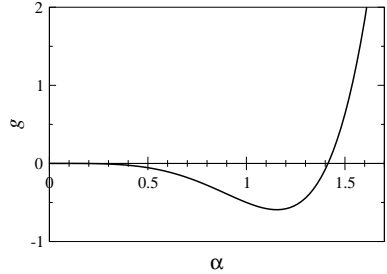}} 
\vspace{.0cm}
\caption{$g(\alpha)$ function, Eq. (\ref{g_vs_alpha}).}
\label{g_plot}
\end{figure}
For $V > V_{+}$, the $V$-dependence of the current is given by $P^0_1(V)$.
With increasing $V$, $P^0_1(V)$ also increases and saturates 
to some quantity (depending on $\alpha$) 
which is comparable with $P^0_0$ (the right-hand side plateau on Fig. \ref{iv_peaks}). 
From numerics we find that at $V=2.5 \Omega$, for instance, for $\alpha = \{ 0.5, 1.1, 1.9 \}$, 
we correspondingly have $P^0_1 \approx 1 - P^0_0 =\{ 0.35, 0.48, 0.38 \}$.
From Eq. (\ref{curr_diff}) and Fig. \ref{g_plot} we see that the NDC plateau disappears 
for small $\alpha < 0.5$; with increasing $\alpha$, the NDC contribution from an open channel
increases-decreases having the maximum at 
$\alpha \approx 1.15$, and for $\alpha > 1.4$ 
the NDC contribution converts to the PDC one (cf. Fig. \ref{iv_peaks}).

Up to now, we have considered only strong asymmetric capacitances and equal capacitances. 
A common feature of these choices is that the current voltage characteristics displays steps
(with or without NDC behavior) whose spacing in voltage is essentially always the same
at relatively low voltages. Indeed, for the asymmetric capacitances plots each step is 
separated by (approximately) $\Omega$, while for $c_l=c_r$, this spacing is $2 \Omega$.
If one was to plot the differential conductance, as in the experiment of 
Ref. \onlinecite{leroy}, one would obtain a periodic series of peaks.
What happens to this periodicity in an intermediate situation where 
the capacitances are comparable, but not equal ? 
\begin{figure}[!!h]
\epsfxsize 7cm
\centerline{\epsffile{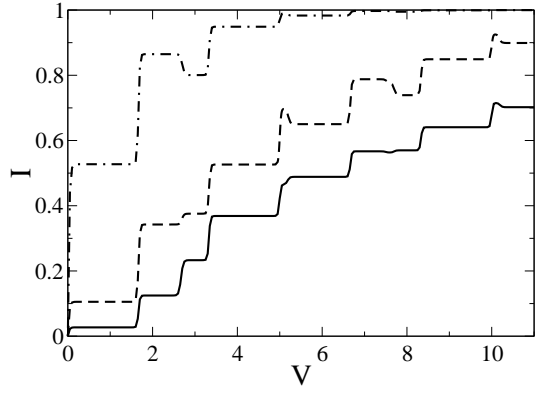}} 
\vspace{.0cm} 
\caption{$\alpha=1.9$ (solid), 1.5 (dashed) and 0.8 (dot-dashed) 
for $c_l=0.4$, $E=0$, $T=0.01$, $s=0$.}
\label{iv_twoperiodic}
\end{figure}
Fig. \ref{iv_twoperiodic} shows 
current-voltage characteristics with two steps of different length for the case 
$c_r = 1.5 c_l$. The number of phonon sidebands above the polaron level, $N$, 
increases by one with the voltage period 
$\Omega/c_r = (5/3) \, \Omega$; for these steps we {\it always} have a PDC behavior
as it is demonstrated in the previous section. 
At the same time, the number of phonon sidebands below the polaron level 
$M \rightarrow M + 1$ with the voltage period $\Omega / c_l = 2.5 \, \Omega$ :
one more channel becomes open which yields the possibility of NDC.
The larger $\alpha$, the more PDC-steps at low bias voltages are observed.

\subsection{Shifted polaron level ($E \neq 0$)}

\begin{figure}[!h]
\epsfxsize 6cm
\centerline{\epsffile{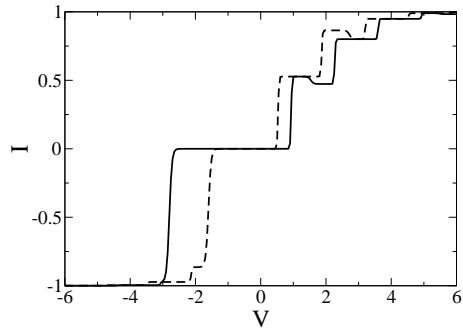}} 
\caption{$E=0.4$ (dashed) and $E=0.7$ (solid)
for $\alpha=0.8$, $T=0.01$, $c_l=0.25$.}
\label{iv_cl025Eshifted}
\end{figure}

In this subsection, 
we consider the more general situation where, due to gate or background charges, 
the polaron level $E$ is shifted by a constant value with respect
to the chemical potentials of the leads at zero bias voltage. 
For $E \neq 0$, the  symmetry $I(V) = - I(-V)$ is violated,
moreover $I(V)$ is not shifted in a trivial manner, meaning that it does not simply follow $E$. 
Fig. \ref{iv_cl025Eshifted} shows two current-voltage characteristics for different values of $E$.
As it was mentioned in Sec. IV, the current is zero when the polaron level lies beyond 
the bias voltage window. For negative $V$, the polaron level becomes captured in this window at
$V = - E / c_l = - 4 E$. At this point, the number of phonon subbands below the bare level
-- $N$ in this case -- suddenly changes from $N=0$ to 
$N = {\rm int} (c_r E / c_l \Omega) = {\rm int} (3 E / \Omega)$. In the curves
of Fig. \ref{iv_cl025Eshifted} corresponding to $E=0.4 \Omega$ and $0.7 \Omega$,
we observe a transition to $N = 1$ and $N=2$, respectively.
This results in a fast saturation of the curves at negative voltages
(the saturation is faster for the curve with larger $E$).

Previously, we found that small $c_l$ leads to PDC for the first phonon assisted peaks.
Here, although $c_l=0.25$ is relatively small in Fig. \ref{iv_cl025Eshifted},
for positive $V$, we observe NDC behavior already for the first phonon-assisted step on $I(V)$
(around $V = 2 \Omega$) for $E=0.7$. This can be explained as follows.
At small positive voltages, we first have a zero-current plateau until  
the polaron level becomes captured
in the bias-voltage window ($N=M=0$) at
$V_0 = E / c_r$. This leads to the first PDC step on the  $I(V)$ curve at 
$V_0 = (4/3) E$. Then the second plateau persists until  
$V_1 = (\Omega - E)/ c_l < 2 \Omega$, which corresponds to   
the phonon sideband level $E-\Omega$ becoming captured in the bias window 
($N=0$, $M=1$). Because $\alpha = 0.8$ is less than the critical value $1$ (see Sec. IV), 
this transition gives the NDC step.
Next a plateau persists until, $V_2 = (\Omega + E)/ c_r$ when the phonon sideband level 
$E+\Omega$ is captured in the bias window ($N=1$, $M=1$) resulting in a PDC step on $I(V)$.
The conditions to observe  NDC for the first phonon-assisted step are therefore:  
\begin{equation}
c_r \Omega > E > (c_r - c_l) \Omega > 0 \;,\;\;\; \alpha < 1 \,.
\label{conditionNDC}
\end{equation} 
The width of the NDC plateau is $V_2 - V_1 = \left( E -  (c_r - c_l) \Omega\right) / c_l c_r$. 
Eq. (\ref{conditionNDC}) is fulfilled for the case $E= 0.7 \Omega$ on Fig. \ref{iv_cl025Eshifted}, 
but it fails for $E = 0.4 \Omega$, where we have the PDC behavior of the first phonon-assisted
step on $I(V)$, for $V > 0$.

\begin{figure}[!h]
\epsfxsize 6cm
\centerline{\epsffile{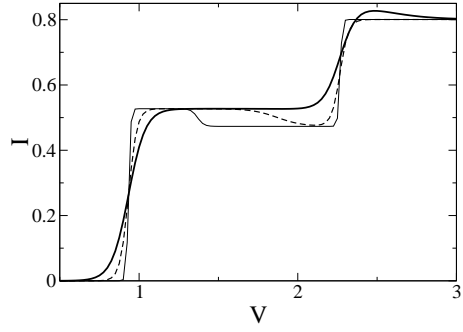}} 
\caption{ The case $\alpha=0.8$, $c_l=0.25$, $E=0.7$ at different temperatures:
T = 0.01 (thin solid), 0.02 (dashed) and 0.04 (bold solid).}
\label{iv_Tdependence}
\end{figure}

Fig. \ref{iv_Tdependence} shows the temperature dependence of the NDC step for $E = 0.7 \Omega$.
With increasing temperature $T$, the phonon-assisted channel opens for higher voltages
when the one-phonon excited state of the empty dot becomes unreachable for thermally activated
electrons from the right electrode at energy $E - \Omega + W$, 
(as before, $W$ is the half-width of the Fermi function).    
As a result, the NDC region with a decreased slope gradually shifts towards the next step on $I(V)$.

\begin{figure}[!!h]
\epsfxsize 5cm
\centerline{\epsffile{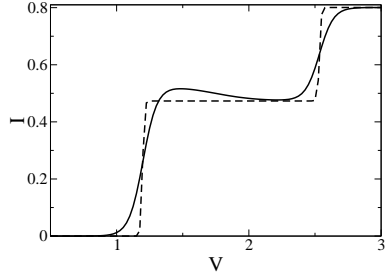}} 
\caption{The NDC induced by temperature
for the case $\alpha=0.8$, $c_l=0.25$, $E=0.9$ : $T=0.005$ (dashed) and 0.04 (solid). 
}
\label{iv_TinducedNDC}
\end{figure}

Fig. \ref{iv_TinducedNDC} shows the opposite situation: the NDC appears this time 
with {\it increasing} temperature. At low temperature, when the Fermi half-width $W$ is negligibly small, 
we only observe PDC behavior
for the first step on $I(V)$ at $V_0 = E / c_r$, where we suddenly have a transition to 
$\{ N=0, M = 1 \}$. 
At this threshold, both the polaron level and the first phonon subband below it 
belong to the bias voltage window. 
Then we obtain a plateau in the $I(V)$ curve up 
to $V_1 = (E + \Omega) / c_r$, where the next PDC step is
associated with the increase of $N$ by one. However, if the temperature is increased so that 
$W$ becomes comparable with $\Omega - E$, now at $V=V_0$
the occupation probability
$P_1^0$ will be negligibly small because of thermally activated electrons tunneling from the right electrode.  
$P_1^0$ starts developing only at $V > (\Omega - E + W )/c_l \approx 1.2 \Omega$ leading to NDC behavior.
Here we have used that at $T = 0.04 \Omega$, 
the half-width of the Fermi function is about $0.2 \Omega$. 
At much higher temperature everything is of course washed out (not shown).

\begin{figure}[!h]
\epsfxsize 7cm
\centerline{\epsffile{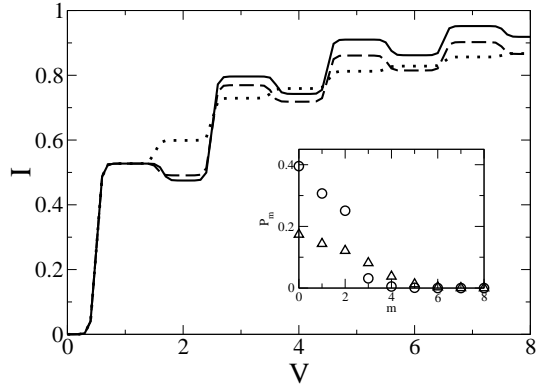}} 
\caption{ For $\alpha=0.8$, $T=0.02$, $E=0.25$, $c_l=0.5$:
$\Gamma_l / \Gamma_r = 10^{-2}$ (solid), 0.1 (dashed) and 1 (dotted).
The inset shows the phonon distribution $P_m^0$ at the bias voltage $V=4$: 
circles for $\Gamma_l / \Gamma_r = 10^{-2}$  and triangles for $\Gamma_l / \Gamma_r = 1$.
}
\label{iv_cl05E025}
\end{figure}

We conclude this subsection by some comments on the role of the asymmetry in the tunneling 
rates $\Gamma$'s on current transport. Fig. \ref{iv_cl05E025} shows $I(V)$ characteristics 
for the different cases of junction asymmetry. 
As it is seen from the plots, the NDC steps in the cases of high and moderate asymmetry 
($\Gamma_l / \Gamma_r = 0.01$ and 0.1, respectively) turn into PDC steps
in the symmetric case ($\Gamma_l = \Gamma_r$). 
We also notice that the NDC steps are slightly shifted with respect 
to their PDC counterparts in the symmetric case. This shift is associated with the finite
width of the Fermi function, and it has been already been discussed in the two previous figures. 
The height of the phonon-assisted steps in the asymmetric case is noticeably large compared 
to the symmetric one. 
This is related to the phonon distribution, an example of which (at $V=4 \Omega$) 
is shown on Fig. \ref{iv_cl05E025} for  both cases (symmetric and asymmetric). 
In the asymmetric case, the number of phonons is restricted by $M$, 
while in the symmetric case the phonon distribution is more spread out.
As a result, in the symmetric case, the phonon-assisted contribution to the current is weakened. 

\subsection{Effect of the half-shuttle on $I(V)$}

In this subsection we discuss the role of the half-shuttle mechanism on $I(V)$ characteristics.

It is worth to be mentioned that in the absence of the half-shuttle ($s = 0$),
the rate equations (\ref{rate_eqs}) are invariant under the transformation
\begin{equation}
V \rightarrow -V \;,\;\;\; E \rightarrow - E \;,\;\;\;
P^0_n \leftrightarrow P^1_n ~.
\label{kineq_symmetry}
\end{equation} 
As a consequence, the current-voltage characteristics $I_E(V)$ and $I_{-E}(V)$ 
corresponding to the cases of the polaron level shifted by $E$ and $-E$, respectively, are related
as follows:
\begin{equation}
I_E(V)=-I_{-E}(-V)~.
\label{iv_symmetry}
\end{equation}
The presence of half-shuttle mechanism ($s \neq 0$) violates this symmetry.

\begin{figure}[!h]
\epsfxsize 7cm
\centerline{\epsffile{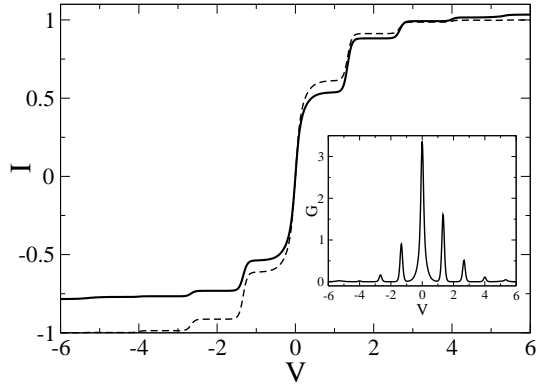}} 
\vspace{.5cm} 
\caption{Asymmetry of $I(V)$ due to "half-shuttle" only: $s=0.1$ (solid) and $s=0$ (dashed).
The case of $\alpha=0.7$, $c_l=0.25$, $E=0$, $T=0.04$. The inset shows the differential conductance
for $s=0.1$.}
\label{shuttle_cl025}
\end{figure}

\begin{figure}[h]
\epsfxsize 7cm
\centerline{\epsffile{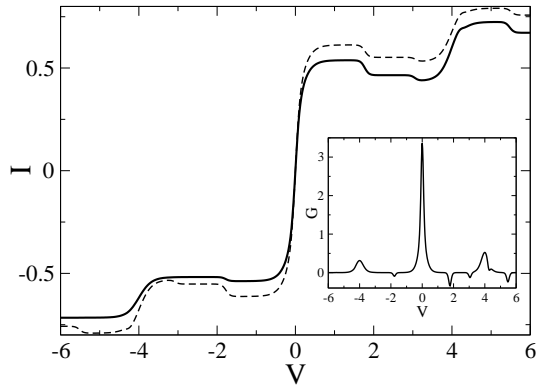}} 
\vspace{.5cm} 
\caption{Asymmetry of $I(V)$ due to "half-shuttle": $s=0.1$ (solid) and $s=0$ (dashed). 
The case of $\alpha=0.7$, $c_l=0.75$, $E=0$, $T=0.04$. The inset shows the differential conductance
for $s=0.1$.}
\label{shuttle_cl075}
\end{figure}

Fig. \ref{shuttle_cl025} shows $I(V)$ for the case when $c_l = 0.25$  is relatively small 
and where the polaron level is not shifted, both of which are favorable for PDC. 
The current steps are suppressed  at negative voltages
(differential conductance peaks in the inset of Fig. \ref{shuttle_cl025}), 
while at positive bias voltage, the current steps have a tendency to increase.
From Eqs. (\ref{I_posV}), (\ref{I_negV}) we obtain that the current-step amplitude at
$V = n \Omega / c_r$ corresponding to the current increase is given by 
\begin{equation}
\Delta I^{(n)} = \Gamma_l \, e^{-\alpha^2 - 2 s \alpha + s^2} \, 
\frac{\left( \, \alpha + {\rm sign}(V) \, s \, \right)^{2 n}}{n!}~.
\end{equation}
In other words, the half-shuttle mechanism works in favor of the formation of the polaron state:
the probability of phonon-assisted tunneling onto the dot from the left electrode is increased,
while the phonon-assisted tunneling from the dot to the left electrode is decreased.
Eq. (\ref{Isat_shuttle}) for the saturation currents illustrates this fact.

The case of relatively large $c_l=0.75$ is shown on Fig. \ref{shuttle_cl075}.
Like in the previous case, here we also observe suppression of the current steps at 
negative voltages. At positive voltages, the NDC steps become more 
pronounced compared to the PDC ones. Expressions for the one-phonon assisted steps are given by 
Eqs. (\ref{deltaI}), (\ref{deltaI_negV}).

\begin{figure}[h]
\epsfxsize 7cm
\centerline{\epsffile{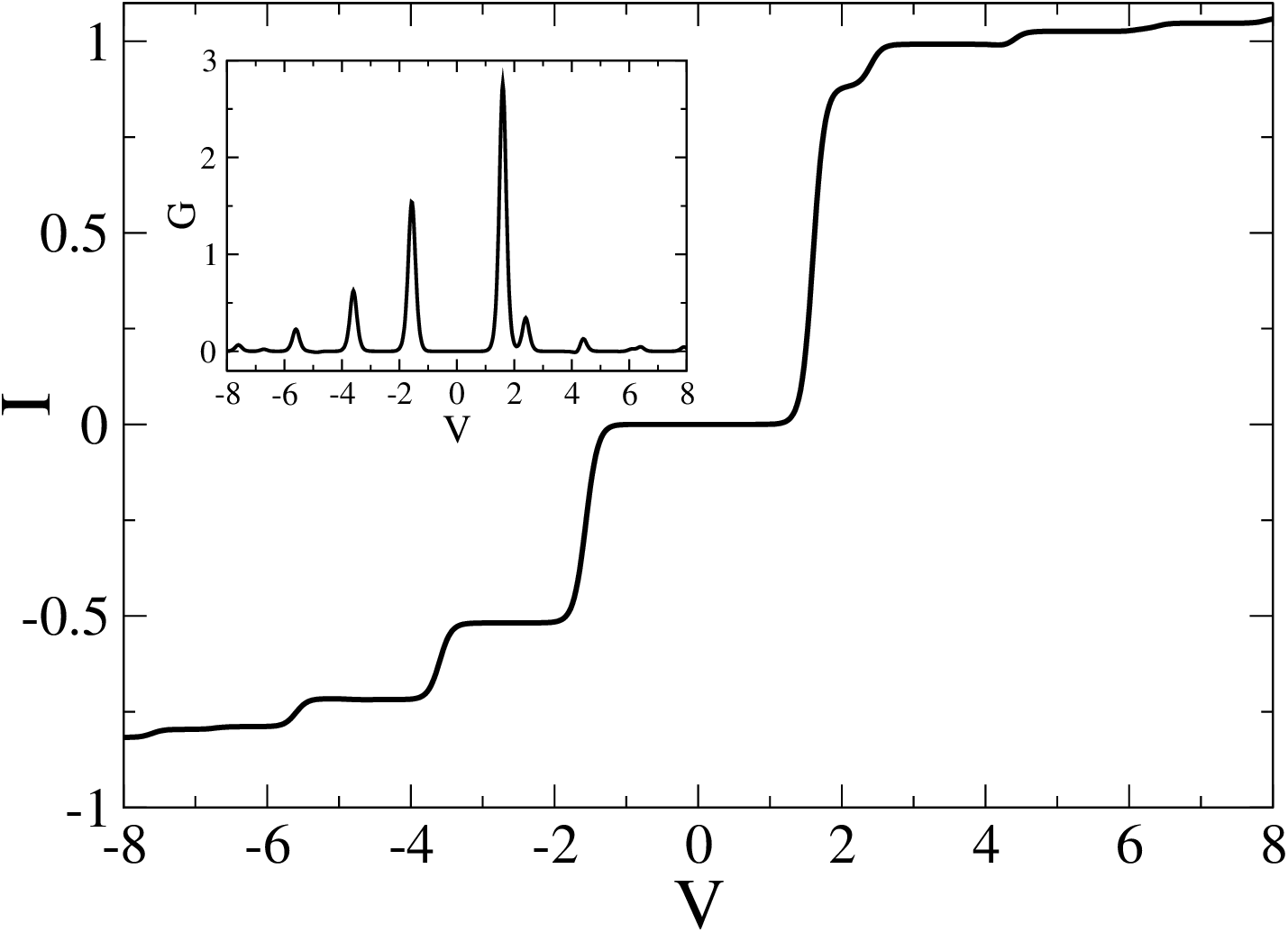}} 
\vspace{.5cm} 
\caption{The case $\alpha=0.7$, $c_l=0.5$, $E=-0.8$, $T=0.04$, $s=0.1$.}
\label{shuttle_Eneg}
\end{figure}
\begin{figure}[h]
\epsfxsize 7cm
\centerline{\epsffile{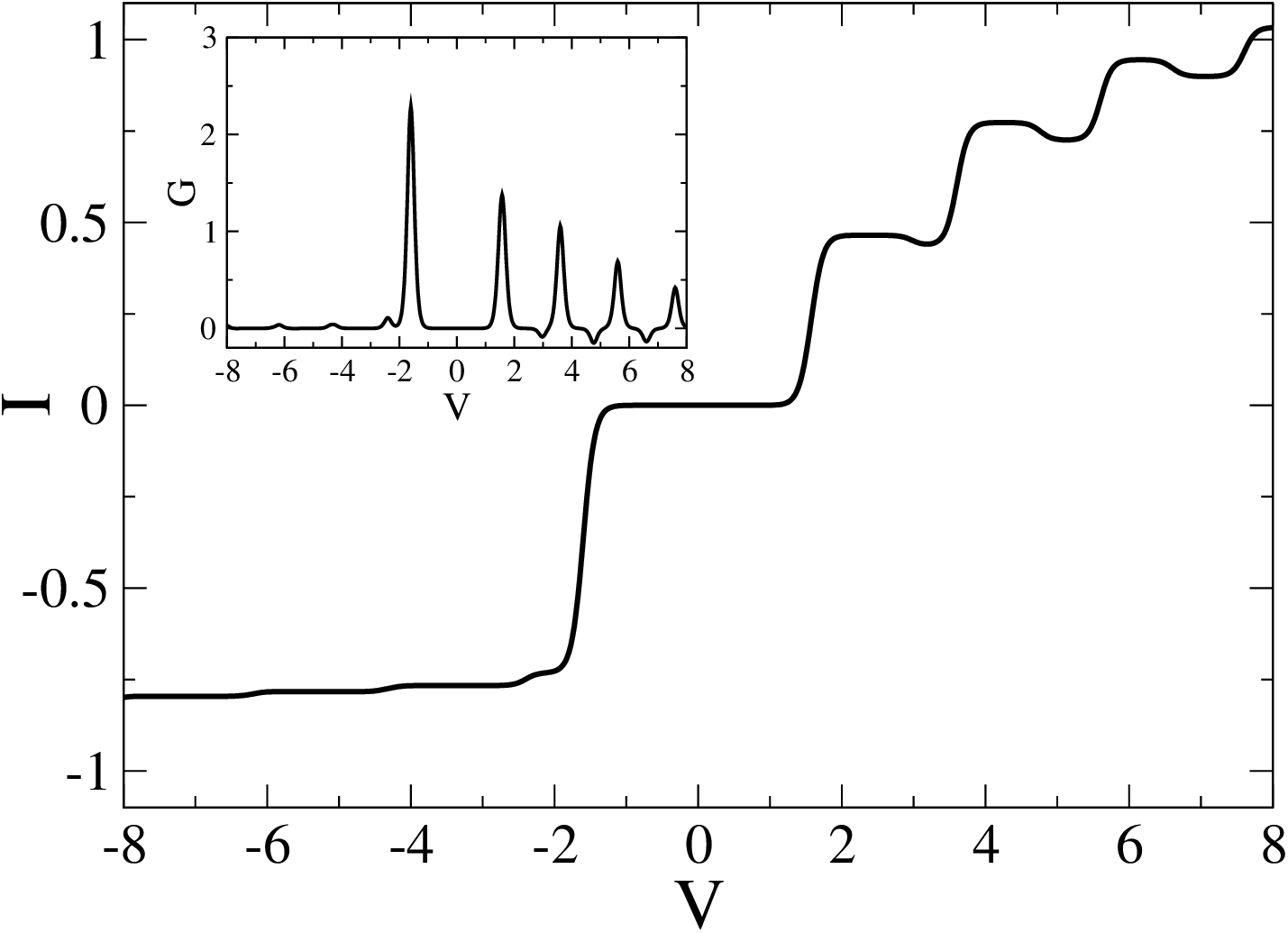}} 
\vspace{.5cm} 
\caption{The case $\alpha=0.7$, $c_l=0.5$, $E=0.8$, $T=0.04$, $s=0.1$.}
\label{shuttle_Epos}
\end{figure}

We have provided concrete evidence that the half-shuttle mechanism
alone, with reasonable values of the parameters, cannot produce NDC 
when $c_l$ is small and when the bare polaron level
is not shifted upwards. Nevertheless, based on the results of Figs. \ref{shuttle_Eneg}
and \ref{shuttle_Epos}, we conclude that in situation when NDC is present
(shifted level or reversed capacitances), the addition of the half-shuttle 
tends to emphasize the NDC features.

\section{conclusion}

To summarize, we have provided an in-depth study of NDC behavior in molecular 
quantum dots or transistors. 
While this study covers a wide range of experimental parameters such as capacitances and 
electron-phonon coupling, here the focus was put on the case where the tunneling 
coupling from the dot to the source and drain leads is asymmetrical. This choice was 
motivated by the experiment of 
Leroy and coworkers\cite{leroy,leroy3} where a nanotube is 
suspended over a trench, allowing nearly free vibrations of the phonon breathing mode. 
Electrons flow from an STM tip to the suspended region of the nanotube, and their tunneling
amplitude is much smaller than that of the nanotube to the substrate. It turns out that this 
assumption allows us to quantify NDC behavior by analytical means. Yet, moderate asymmetries, 
often encountered with two poor metallic contacts, qualitatively display the same physics.
        
We have used a microscopic approach modeling the molecule/nanotube as a quantum dot, 
which is justified based on the early observation of Coulomb blockade behavior in carbon 
nanotubes. A single phonon mode couples on-site to the quantum dot, in two different way.
First, as the principal mechanism, it couples to the electron density on the dot, 
as in a polaron model. Second, the 
half-shuttle coupling was introduced, with the motivation that it should be present in 
the experiment of Refs. \onlinecite{leroy,leroy3}: when the nanotube vibrates, the tunneling 
distance between the nanotube wall and the STM tip oscillates accordingly, but the 
amplitude of this motion is believed to be rather small (or the order of the zero point 
motion, which can be smaller than $0.1\AA$). 

With these ingredients, the polaron transformation eliminates 
the electron charge coupling to the phonon on the dot, and transfers this coupling
to the tunneling Hamiltonian. At this point, we have chosen to describe the situation
corresponding once again to the experiment of Ref. \onlinecite{leroy,leroy3}, 
where the time scale associated with tunneling events between the dot and the
leads is large compared to the temperature, so that electrons evacuated in the 
leads effectively loose their phase coherence. On the other hand, the quantum mechanical 
nature of electron-phonon dynamics within the dot (polaron formation) is fully captured. 
This happens to be the regime were most molecular electronics experiments 
are performed nowadays, as these experiments typically do not require dilution 
refrigerator technology. From the density matrix of the total system, a kinetic 
equation for the reduced density matrix (with the leads degrees of freedom integrated
out) was derived, leading to master equations for the electron population
and their associated phonon numbers. 

At temperature high enough for this rate equation 
is valid, but low enough that thermally activated tunneling from the phonon sideband
can be neglected, analytical predictions were made for the current-voltage characteristics, 
thanks to our assumption of a highly asymmetrical molecule--lead configuration: 
the positive or negative differential conductance behavior depends both on the location
of the bare polaron level and on the occupation of the associated phonon sidebands
-- above and below this level -- which are included in the bias voltage window. 

Turning to numerics, attention was first drawn to the role of the relative capacitances 
to the dot. If the molecule/nanotube is in better contact with the lead which 
evacuates the electrons (the substrate), than with the one who injects electrons, 
it is plausible to believe that the capacitance of this junction will be larger
than that of the injecting junction. In this case, no negative differential conductance
is obtained. When increasing the bias voltage, the contribution 
of an increasing number of phonon sidebands leads to a standard staircase behavior in the 
current-voltage characteristics. Thermal effects tend to round off the first step.  
In the experiment of Ref. \onlinecite{leroy,leroy3}, the use of several nanotube samples with 
different lengths allowed to vary the relative capacitance of the injecting 
and of the tunneling contact. This provided a motivation to study the so-called
``reversed capacitance'' case, which according to our analytical prediction
could justify the presence of an negative differential behavior. This was indeed 
observed numerically, but it was emphasized that intermediate phonon couplings
are needed to observe NDC: ``small'' electron-phonon coupling leads to barely
noticeable NDC, while ``large'' electron-phonon coupling suppresses NDC altogether, 
thus the need to use non-perturbative techniques to study NDC. 

Do these results disqualify the possibility of NDC if contacts where the capacitances 
between left and right are comparable ? The answer is no, although NDC could be more 
difficult to detect in this case. While the global shape of the current voltage 
characteristics displays the staircase structure typically attributed to PDC, the 
onset of each step is followed by a small spike, thus displaying NDC behavior. 
The amplitude of this peak can be optimized with phonon coupling. 
For slightly asymmetric capacitances, the same effects are observed, but the 
staircase structure is modified as it involves two distinct periodicities. 

Because background charges may shift the bare polaron level, and because of the possibility
of using a voltage gate on the molecule/dot we also studied the
possibility of NDC in this case. This suggests that 
when the level is shifted upwards, NDC is possible even when the capacitance 
of the injecting junction is small.

Finally, we have proposed that a half-shuttle mechanism may play a role in a STM 
experiment. It can be detected by the asymmetry of the current-voltage curves. 
Yet, in our opinion, this mechanism cannot account for NDC alone and only tends to increase the heights 
of PDC and NDC steps in the current-voltage characteristics. Notice that approaching the tip closer
to the molecule should strongly enhance the half-shuttle mechanism.   
In our numerical calculations, we set $s=0.1$ assuming 
the amplitude of the zero-point RBM oscillations to be of the order of $1$pm.
Estimating the zero-point amplitude, we have considered a single-wall carbon nanotube
as an elastic hollow cylinder with diameter $d_0 \sim 1$ nm, with wall thickness $\sim 1~\AA$,
and with a length corresponding to the relevant breathing part $\sim 10$ nm. 
Accordingly, the undeformed volume of the cylinder is $V_0 \sim 1 {\rm nm}^3$. 
The elastic properties of the nanotube cylinder are characterized with 
the average Young's modulus $Y \sim 1$ TPa (see, for instance, Ref. \onlinecite{popov}).
Assuming $\hbar \Omega \approx 10$ meV ($\Omega$ is the frequency of the RBM), 
one finds for the relative extension of the nanotube due to the breathing, 
\begin{equation}
\delta x \sim \left( \frac{\hbar\Omega}{Y~ V_0} \right)^{1/2} \sim 10^{-3}~.
\end{equation}
Thus, the estimated diameter change, $\sim \delta x ~ d_0$, 
associated with the zero-point oscillations is of the order of 1 pm.

Overall, we have provided a rather complete account on the possible occurrence of
NDC behavior in molecular electronic transport, due to phonon-assisted transport. 
This phenomenon is due to a distribution of the total spectral weight of the injected electron on 
the polaron levels included in the bias window. Taking into account the correct microscopic 
expressions of the Franck-Condon factors associated to phonon transitions, we have demonstrated 
that NDC effects are a fingerprint of the polaronic nature of charge carriers. 
While we cannot rule out that NDC could 
exist in molecular system with symmetric molecular junction, it was not observed 
within this model. On the other hand, we have clearly shown that for 
asymmetric tunneling rates, there exist a wide range of parameters
which lead to NDC.    
      
A striking result of this work is the fact that a simple counting of the NDC steps
in the current voltage characteristic (or alternatively, a counting of the peaks
in differential conductance) provides direct information about how many phonon
subbands contribute to transport.              
The question remains about whether this NDC behavior can be observed in experiments. 
Our model could apply to Ref. \onlinecite{sapmaz}, if the tunnel couplings are indeed asymmetric.
Concerning Ref. \onlinecite{leroy}, NDC occurs but only for the first phonon sideband. 
This can be due 
to several factors : first, as a general rule, the height of PDC or NDC steps  
tends to decrease with increasing phonon sidebands number. Furthermore, if the 
electron phonon coupling is weak, NDC is present but its manifestation is weak. 
Finite temperature tends to smear these steps, so in order to observe several 
peaks in NDC, lower temperatures would be required, keeping in mind that the 
tunneling contacts should not be too large in order to ``avoid'' any Kondo
like regime. Another aspect which we have not considered in this work is 
the possibility of phonon damping. The population of the phonon subbands
may rearrange due to the coupling with the environment of the molecule \cite{flensberg,aleiner}. 
The environment could include electron-hole pairs generated in the leads, 
which should be reduced because of our weak coupling assumption. On the other 
hand, for a ``large'' molecular system, the optical phonon mode (the breathing 
mode of the nanotube of Ref. \onlinecite{leroy}) is also accompanied by other 
phonon modes such as acoustic-like modes, which have typically lower energies, 
and which therefore can play the role of a ``bath''. Because of the importance 
of the occupation of the phonon subbands for NDC in our present work, we expect 
that substantial coupling to this bath will tend to suppress NDC.              
   
\acknowledgements

Discussions with V. Mujica and H. van der Zant are gratefully acknowledged.


\end{document}